%% file: Main.tex
\documentclass[sigconf,nonacm]{acmart}

\setcopyright{none}
\usepackage[normalem]{ulem}
\usepackage{float}
\usepackage{multirow}
\usepackage{enumitem}
\usepackage{longtable}
\usepackage{csquotes}
\usepackage{colortbl}
\usepackage{float}
\usepackage{placeins}
\usepackage{indentfirst}
\usepackage{hyperref}
\usepackage{makecell}
\usepackage{soul}
\usepackage{listings}
\usepackage{textcomp}
\usepackage{tabularx}
\usepackage{diagbox}
\usepackage{pifont} 
\usepackage{booktabs}
\usepackage[normalem]{ulem}
\usepackage{graphicx}
\usepackage{array}
\usepackage{wrapfig}
\usepackage{capt-of}
\usepackage{longtable}
\newlength{\stimulusTableWidth}

\definecolor{black}{RGB}{0,0,0}
\definecolor{blue}{HTML}{005CFF}
\definecolor{grey}{HTML}{666666}

\AtBeginDocument{%
  \providecommand\BibTeX{{%
    \normalfont B\kern-0.5em{\scshape i\kern-0.25em b}\kern-0.8em\TeX}}}

\copyrightyear{2027}
\acmYear{2027}
\setcopyright{cc}
\setcctype{by}
\acmConference[]{}{}{}
\acmBooktitle{}
\acmPrice{}
\acmDOI{}
\acmISBN{}

\begin{document}
\raggedbottom

\title[Crediting as a Trust Signal for AI-Generated Content]{Unspectacular Feed: Crediting as a Trust Signal for AI-Generated Content in Social Media Feeds}

\author{Saetbyeol LeeYouk}
\email{sbleeyuk@mit.edu}
\affiliation{%
  \institution{MIT Media Lab}
  \city{Cambridge}
  \state{Massachusetts}
  \country{United States of America}
}

\author{Karrie Karahalios}
\email{kkarahal@mit.edu }
\affiliation{%
  \institution{MIT Media Lab}
  \city{Cambridge}
  \state{Massachusetts}
  \country{United States of America}
}

\renewcommand{\shortauthors}{LeeYouk et al.}

\begin{abstract}
“Is this AI?” Photorealistic generative media complicates credibility judgments. Binary AI labels provide limited information about how, where, and why AI was used, leaving viewers to distinguish simple editing from misleading generation. Our formative study identified creators’ needs for crediting varied AI uses and viewers’ needs for production context beyond binary labels. We present \textit{Unspectacular Feed} making different degrees of AI involvement inspectable through process-based crediting, creator disclosure histories, and baseline AI-image detection. Drawing on social signaling theory and journalistic principles of source transparency, the system presents production records and disclosure histories as contestable credibility signals. Creators disclose source images, prompt histories, purposes, and generated regions; viewers inspect these alongside detection results and can filter generated regions. In a controlled study with 15 creators and 15 viewers examining AI-generated conflict-related content, creators reported that disclosures helped them communicate production decisions to audiences precisely. Viewers used disclosures and credit history to assess trustworthiness and reported greater awareness of partial AI editing cases beyond fully generated images. These findings suggest that contextual disclosure can support informed trust judgments and help viewers develop a more nuanced understanding of AI involvement, while motivating future research on disclosure incentives and detection uncertainty.
\end{abstract}



\maketitle

\input{V2/01_Introduction}
\input{V2/02_Related_Works}
\input{V2/03_Formative_Study}
\input{V2/04_Design_Goals}
\input{V2/05_Interface}
\input{V3/06_User_Study}
\input{V2/07_Discussion}
\input{V1/08_Limitation}

\input{V1/09_Conclusion}
\input{V2/10_Gen_AI}


\bibliographystyle{ACM-Reference-Format}
\bibliography{References}
\clearpage

\input{V2/10_Appendix}

\end{document}

%% file: V2/01_Introduction.tex
\section{Introduction}
With the rise of consumer-facing generative AI products, photorealistic images and videos have become inexpensive to produce and circulate at scale, reshaping online media consumption. This proliferation of AI-generated contents (AIGC) has raised concerns about misinformation \cite{10.1145/3706599.3719834, dufour2024ammeba, epstein2023label}, polarization of public discourse, and declining trust in mediated evidence \cite{ibrahim2024swift, ccdh2024factories}. Limited accountability for the production and circulation of fabricated imagery further complicates how viewers interpret events and assign responsibility \cite{kahn2023factcheckers}. Especially, in high-stakes contexts such as war and natural disasters, increasingly spectacular synthetic depictions can heighten expectations of what suffering and destruction should look like, leading viewers to underestimate real events when authentic documentation appears less extraordinary. Meanwhile, awareness of AI fabrication enables authentic evidence to be dismissed with claims that it “looks like AI.” This vulnerability, known as the liar’s dividend, allows actors to exploit doubts about authenticity to evade accountability \cite{chesney2019deepfakes}. These dynamics weaken the evidentiary role of visual media and amplify the influence of emotional intensity on how events are evaluated.

These concerns have prompted regulatory and platform-level efforts to make AI involvement visible\cite{euaiact2024, californiaSB942}. Following this, platforms have introduced disclosure mechanisms, including AI-origin information in Google Search and Meta’s “AI info” labels \cite{meta2024labeling, adobeStockGenAI, youtubeAIdisclosure, linkedinCredentials}. These mechanisms identify binary AI involvement, but labels alone provide limited information about the scope, purpose, and sequence of interventions behind a post.

Journalism ethics offers a complementary distinction between documentary evidence and illustration. The International Federation of Journalists' Global Charter emphasizes verification and respect for facts, while the Society of Professional Journalists' Code of Ethics calls for clearly labeling illustrations and reenactments \cite{ifj2019charter,spj2014ethics}. This distinction is useful for thinking about \textit{\textbf{enactment}} on social media: AI-generated or AI-edited imagery can help viewers imagine aspects of an event when documentary imagery is unavailable, but it can also be mistaken for a record of what happened.

In this context, prior work has examined how AI labels affect credibility and sharing, but less is known about what motivates creators to explain their AI use \cite{wittenberg2024promises, weiner2023regulating, wittenberg2024promises}. When viewers read process labels as warnings that content is misleading, creators may fear reputational costs and withhold disclosure \cite{schilke2025transparency, altay2024skeptical, zhang2025secret}. Drawing on journalism ethics, we frame contextual disclosure as a way to help viewers assess how and why AI was used and recognize responsible use. 

We conducted a formative study with two creators and four social media viewers to understand how creators explain their AI use and how viewers interpret those explanations. The study informed our design of contextual disclosures, inspectable production histories, and explicit communication of detection limits. Based on these findings, we introduce \textit{Unspectacular Feed}, which connects process-based crediting, automated detection, and creator disclosure histories to support labeling as a contestable credibility signal. Building on research documenting public concerns about AI's effects on journalism \cite{lipka2025ainews} and audience preferences for disclosures that communicate human oversight and collaboration \cite{zier2026byline}, the system accommodates AI involvement across techniques and stages of production, including interventions that creators may regard as routine editing. It connects harm-oriented warnings \cite{wittenberg2025labeling} with active process disclosure as a basis for trust through three mechanisms: (1) contextual disclosures identifying techniques, purposes, source materials, and affected image regions; (2) baseline AI-image detection that cross-checks declared involvement and makes its assessment available for scrutiny; and (3) Unspectacularized mode, which blurs generated regions and overlays the creator’s disclosed context. Viewers can choose whether to reveal these regions and retain disclosure context when sharing. Post-level records and accumulated creator histories support assessments of both content and uploader trustworthiness. This cycle is designed to make accountable AI use visible and reputationally valuable while giving viewers control over how they encounter and circulate AI involved contents.

We conducted two studies ($N=15$ each) to examine how creators disclose AI use and how viewers assess posts, following hypothetical scenarios that have been used to examine fairness and ethical judgments and users' responses to algorithmic systems~\cite{blader2003procedural,huppertz1978equity,kimes1994fairness,petrinovich1993moral,kray2002injustices, vaccaro2020contesting}. Both studies used US--Iran conflict-related content.\footnote{During the study, the United States was engaged in military action with Iran.} Creators uploaded 3 posts containing AI-generated or AI-edited images, while viewers assessed 9 posts with varying disclosures. Creators used disclosures to explain production choices and flag uncertainty about sources. Viewers consulted disclosures and credit histories, but favorable assessments of disclosure did not consistently translate into trust in the depicted content. Automated detection offered additional information, though participants questioned its reliability. Participants also reflected on production choices and noticing subtler AI edits. These findings suggest that contextual disclosure can support scrutiny of AI use, while highlighting the need to examine the consistency of disclosure records and communicate detection limitations.

This work makes three contributions. Through a formative study, we surface tensions between creators’ understandings of AI use and viewers’ disclosure expectations, informing a positive-disclosure approach that makes production context and creator contributions visible. We present a process-based AI-disclosure system connecting contextual disclosures, automated detection, and creator reputation to support viewer agency in assessing and filtering AI-mediated content. With a user study of 15 creators and 15 viewers, we examine how process disclosures, detection results, and credit histories inform disclosure practices, credibility judgments, and decisions about exposure to AIGC.


%% file: V2/02_Related_Works.tex
\section{Related Works}
\subsection{Social Media Platforms' Approaches to AI Disclosure}
Regulatory frameworks shape platforms' AI disclosure policies. The EU AI Act requires machine-readable marking of synthetic outputs and disclosure of deepfakes, exempting certain standard edits from marking and adapting disclosure requirements for evidently artistic, fictional, or satirical works~\cite{euaiact2024}. In the United States, the proposed AI Labeling Act of 2023 sought clear, conspicuous, and technically persistent disclosures on generated content, with audiovisual metadata identifying the generating tool and creation time~\cite{ailabelingact2023}. California's enacted AI Transparency Act requires covered providers to embed latent disclosures and offer visible disclosure options for generated audiovisual content~\cite{californiaSB942}.

Platforms combine uploader declarations with automatic labeling. TikTok and YouTube require disclosure of realistic AIGC or meaningfully altered content and label outputs from their platform built-in AI tools~\cite{tiktokAIGC,youtubeAIdisclosure}. Meta uses self-disclosure and technical signals, while LinkedIn displays available Content Credentials~\cite{meta2024labeling,linkedinCredentials}. Methods including C2PA credentials contain signed assertions that support provenance verification without establishing factual accuracy~\cite{c2paExplainer}. Watermarks such as SynthID embed imperceptible signals within generated images for identification~\cite{gowal2023synthid}.

These approaches face limitations in coverage, persistence, and interpretation. Screenshots can detach images from embedded metadata, while manipulation can undermine watermark detection~\cite{caiFAQ,gowal2023synthid}. In a recent audit of Instagram, TikTok, X, and YouTube, \citet{rijsbosch2026slop} found that only 61\% of controlled uploads carrying standard AI-provenance signals received labels, while platforms commonly stripped those signals after upload; only 33\% of expert-identified deepfakes in their sampled systemic-risk contexts carried platform-applied labels. Label absence consequently provides insufficient grounds for concluding that content is free of AI involvement. The wording of labeling can obscure the extent of AI usage. Platforms differ in what their labels reveal. TikTok identifies AIGC, Meta explains whether labels arise from technical signals or self-disclosure, and LinkedIn exposes available provenance fields~\cite{tiktokAIGC,meta2024labeling,linkedinCredentials}. 

\subsection{Detection and Forensic Analysis of AIGC}
Automated forensics generalization across generators and robustness to image transformations remain a challenge. Wang et al. demonstrated transfer across CNN generators~\cite{wang2020cnndetection}, and subsequent approaches improved generalization through frozen CLIP representations~\cite{ojha2023universal}, and forensic microstructure representations, supporting zero-shot detection and open-set attribution~\cite{nguyen2025selfdescriptions}. Emerging generators also create a need for adaptation over time. \citet{li2023cddb} CDDB addresses the challenge of learning new sources while retaining previous capabilities. These advances reduce dependence on known generators, but social-media transformations such as compression and resizing can affect detection performance~\cite{corvi2023diffusion}. Domain-specific evaluation is also necessary. A recent satellite-image benchmark preprint identifies limitations in existing datasets' realism, localization masks, and acquisition metadata, highlighting the need to evaluate detection under domain-specific conditions~\cite{arndt2026satellite}.

Interpretability presents a further challenge. FakeShield combines forgery detection with localized masks and textual explanations, making the basis of model judgments available for inspection~\cite{xu2025fakeshield}. Such support is important because unaided human judgment is often unreliable. Tahir et al. documented perceptual difficulties in identifying deepfake videos~\cite{tahir2021seeing}, while Diel et al.'s meta-analysis found detection performance not significantly above chance across the evaluated modalities, with improvements from some assistance and training strategies~\cite{diel2024human}. These findings motivate interfaces that clearly communicate the basis and limitations of detection evidence and its limitations while supporting users' interpretation and critical evaluation.

\subsection{AI Labeling Approaches and Effects}
Research on AI labeling addresses three related questions: what information labels communicate, how that information is presented, and how labels affect credibility judgments and engagement.

\textbf{Label content.} Wittenberg et al. distinguish process-based labels, which describe how content was produced, from harm-based labels, which indicate the potential to mislead\cite{bhuiyan2018feedreflect, weiner2023regulating, wittenberg2024promises}. Both reduced belief in misleading posts, but standalone AI-origin labels had little effect on engagement intentions; harm-based or hybrid labels were better suited to discouraging misinformation sharing~\cite{wittenberg2024promises}. The stated purpose of AI use also matters. Li and Lu found that privacy-protection explanations elicited more favorable responses to generated news images than cost-saving explanations or simple disclaimers, although generated images still reduced organizational trust relative to photographs~\cite{li2026legitimacy}, showing the limitation of a single `AI label' without context of production.

\textbf{Presentation and detail.} Signaling theory distinguishes conventional signals, such as self-declared AI labels whose reliability depends on social enforcement, from handicap signals, whose costs can make deception disadvantageous~\cite{donath2007signals,grafen1990handicaps}. AI labels introduce credibility cues whose design affects trust in labels, although label presence did not significantly change reported engagement~\cite{10.1145/3706598.3713171}. Detailed news disclosures covering AI use, human oversight, accountability, and transparency principles increased source-checking but reduced reported trust and subscriptions, despite participants' preference for detail~\cite{prajod2026disclosure}. These findings suggest that the amount and presentation of disclosure can shape evaluation in ways that are not captured by engagement or trust alone.

\textbf{Credibility and unintended effects.} Disclosure can impose reputational costs. Across 13 experiments, Schilke and Reimann found that disclosing AI use reduced trust in the disclosing actor under both voluntary and mandatory disclosure conditions~\cite{schilke2025transparency}. \citet{altay2024skeptical} linked skepticism toward AI-labeled headlines to assumptions of full automation~\cite{altay2024skeptical}. \citet{holtervennhoff2026labels} found that labels reduced belief in false claims illustrated with generated images but also encouraged overreliance: participants became more receptive to false claims illustrated with human-made images and more skeptical of true claims illustrated with labeled generated images. These results show that provenance labels can influence judgments beyond what they establish about factual accuracy.

These studies examine labels and accompanying explanations, leaving less understood how disclosure can be connected to specific AI-altered regions within visual media. Much of this work focuses on how viewers interpret labels, with less attention to creators' disclosure practices. Further research should examine tensions between the production details creators consider important to disclose and those viewers use to assess credibility.

%% file: V2/03_Formative_Study.tex
\section{Formative Study}
We conducted a formative study with digital content creators and social media viewers to examine creators' AI use and disclosure practices, alongside how viewers perceive, interpret, and evaluate AIGC to identify meaningful signals that communicate creator intent and support trustworthiness judgments.

\subsection{Participants and Procedure}
We recruited two digital content creators (P3, P6) and four social media viewers (P1, P2, P4, P5) through posts on institutional online forums and researcher's social media post (Appendix \ref{app:formative-participants}). P3 was a new media artist who published on Instagram and reported occasionally using generative AI for planning and drafting, but not for final outputs. P6 produced finance and economics content for an Instagram zine affiliated with a Korean broadcaster and used generative AI for still-image generation. Viewers’ reported platform use spanned YouTube, Instagram, Facebook, and LinkedIn. All viewers encountered news or informational content in their feeds daily and had previously encountered content they knew or suspected was AI-generated.

The study comprised three parts (see Appendix \ref{app:formative-disclosure}):
\textbf{(1) Goals of production or viewing practices}. Creators discussed their own content, production goals, and experiences with generative AI; viewers discussed their social media use and experiences encountering AIGC. \textbf{(2) Disclosure expectations.} Following scenario-based approaches in HCI~\cite{vaccaro2020contesting}, participants assumed the role of a news editor and drafted disclosures for three AI-generated images. They then evaluated 15 AI techniques, selecting accept, reject, or uncertain for use in their work (creators) or appearance in their feed (viewers), and explained their disclosure.  \textbf{(3) Prototype feedback.} Participants compared two early prototypes---AI-detection-based flagging and creator-authored production records---and offered design suggestions.

\subsection{Findings}
\subsubsection{\textbf{Tensions between Production Practices and Anticipated Audience Judgments}}
P3 confined AI use to brainstorming and early sketch of idea, while P6 used generative fill to adapt magazine images to layout requirements. P6 had not disclosed these edits, viewing them as minor adjustments rather than AI use requiring disclosure, and worried that an AI label would prompt viewers to scroll past, assuming entire post was AI-generated. P3 recounted backlash against another artist whose animation used AI for post-processing for color calibration, arguing that a general AI label overshadowed the artist’s broader contribution. This tension shows that creators' needs on clarifying the different degree of the AI involvement. 

\subsubsection{\textbf{Acceptance Reflected Perceived Purpose and Human Effort}}
Viewers valued AI use that expanded creative possibilities and conveyed effort. P2 and P5 highlighted filmmaking enabled by limited resources; P4 enjoyed clearly synthetic, playful videos despite concerns about livelihoods; and P5 valued substantive AI-assisted writing, reporting that explicit AI labeling did not undermine trust.
For news and informational content, concerns centered on fabrication and its consequences for trust. P1 and P2 worried that political fabrications could reinforce existing beliefs, while suspicion of fabrication could weaken responses to real events. P4 described difficulty correcting a parent who shared fabricated footage and increasingly presumed content was synthetic. These accounts suggest that viewers assessed AI use through its purpose, the creator's contribution, and the potential harm in its context.

\subsubsection{\textbf{Expectations and Competing Incentives of Disclosure}}
In the three-image news-editor task, disclosure expectations reflected realism and editorial responsibility. P3 distinguished photorealistic fabrication from illustration, P1 and P2 preferred human illustrators, and P5 proposed stronger labels for realistic images. P6 instead proposed a large faint watermark for a war image to capture viewers' attention. This extends evidence of disclosure-related trust penalties~\cite{schilke2025transparency} by showing how anticipated audience reactions can shape disclosure visibility, even where prior experiments found no significant effect of label presence on reported engagement~\cite{10.1145/3706598.3713171}.
Across the 15 AI techniques, participants judged alterations through their purpose and informational consequences. P4 distinguished art from news, P5 distinguished technical adjustments from fabricated information, and P6 reconsidered exempting routine edits after discussing misuse. P3 favored supplementary disclosure to preserve storytelling. These findings show that disclosure thresholds depended on context and remained open to revision.
Prompt disclosure introduced competing priorities: P3 regarded prompts as proprietary creative work, P2 feared misuse, and P5 valued an inspectable record of intent. This complements research identifying prompts as creative work~\cite{chang2023promptartists}, while revealing a tension between acknowledging AI use and exposing its detailed process, motivating options to attach generation histories or summarize prompts.

\subsubsection{\textbf{Feedback on AI Detection and Human-Made Crediting}}
Participants valued granular detection signals but questioned their reliability. P4 welcomed detail beyond binary labels, while P1, P2, and P4 requested evidence of model credibility. P3 anticipated regeneration to evade detection, and P4 and P5 questioned robustness to screenshots, cropping, and reposting, motivating explicit communication of detection limitations. Human-made crediting raised concerns about ambiguous categories and strategic disclosure. These findings motivate disclosure that accommodates changing practices and audience preferences.

%% file: V2/04_Design_Goals.tex
\section{DESIGN GOALS}
Our formative study informs four design goals for a positive-disclosure cycle that encourages creators to disclose varied forms and degrees of AI use without fear of stigma. We connect two trust mechanisms: reputation, expressed through disclosure histories, and proof, understood as inspectable production records and evidence with explicit limitations. 
\textbf{DG1. Make Disclosure Specific, Manageable, and Accountable.} The system supports structured disclosures of where, how, and why AI was used, distinguishing modalities and partial edits without overburdening creators. 
\textbf{DG2. Represent AI Involvement as a Process History.} To reflect creator's iterative AI use process, the system makes this legible through records of prompts, revisions, and successive edits, so viewers can understand intent of developing process.
\textbf{DG3. Make Detection Performance and Uncertainty Transparent.} To communicate the trustworthiness of the detection system, we disclose the detector’s measured accuracy, error rates, evaluation conditions, and known failure modes, so viewers can assess the strength and weakness of detection evidence. 
\textbf{DG4. Connect Disclosure Histories to Creator Reputation and Viewer Choice.} Disclosure creates reputational value for creators while giving viewers information to assess their trustworthiness and act on those judgments. 

%% file: V2/05_Interface.tex
\section{Unspectacular Feed: Implementations}

\begin{figure*}[t]
    \centering
    \begin{minipage}[t]{0.72\textwidth}
        \centering
        \includegraphics[width=\linewidth]{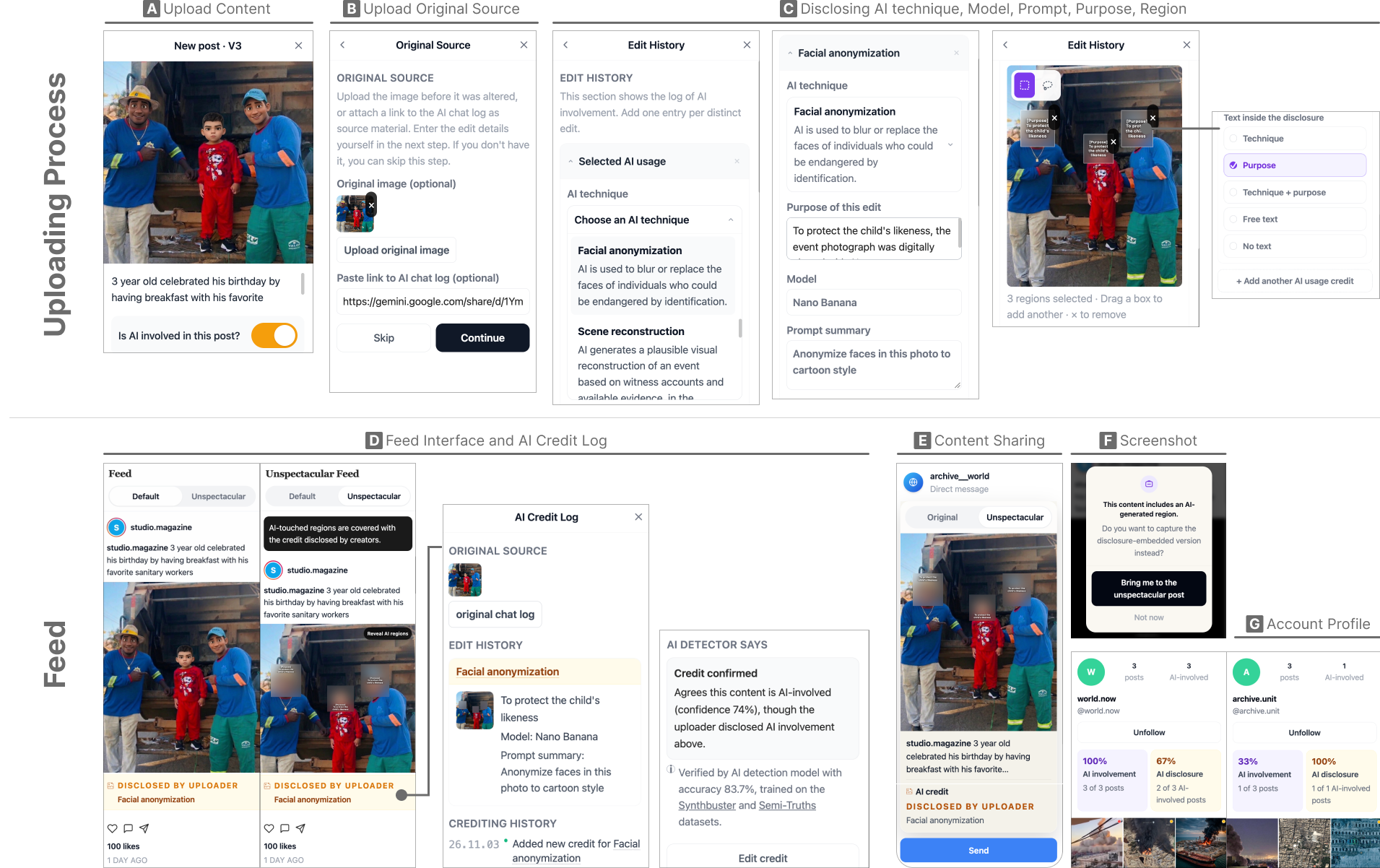}
        \caption{Unspectacular Feed Interface.}
        \label{fig:interface}
    \end{minipage}
    \hfill
    \begin{minipage}[t]{0.25\textwidth}
        \centering
        \includegraphics[width=\linewidth]{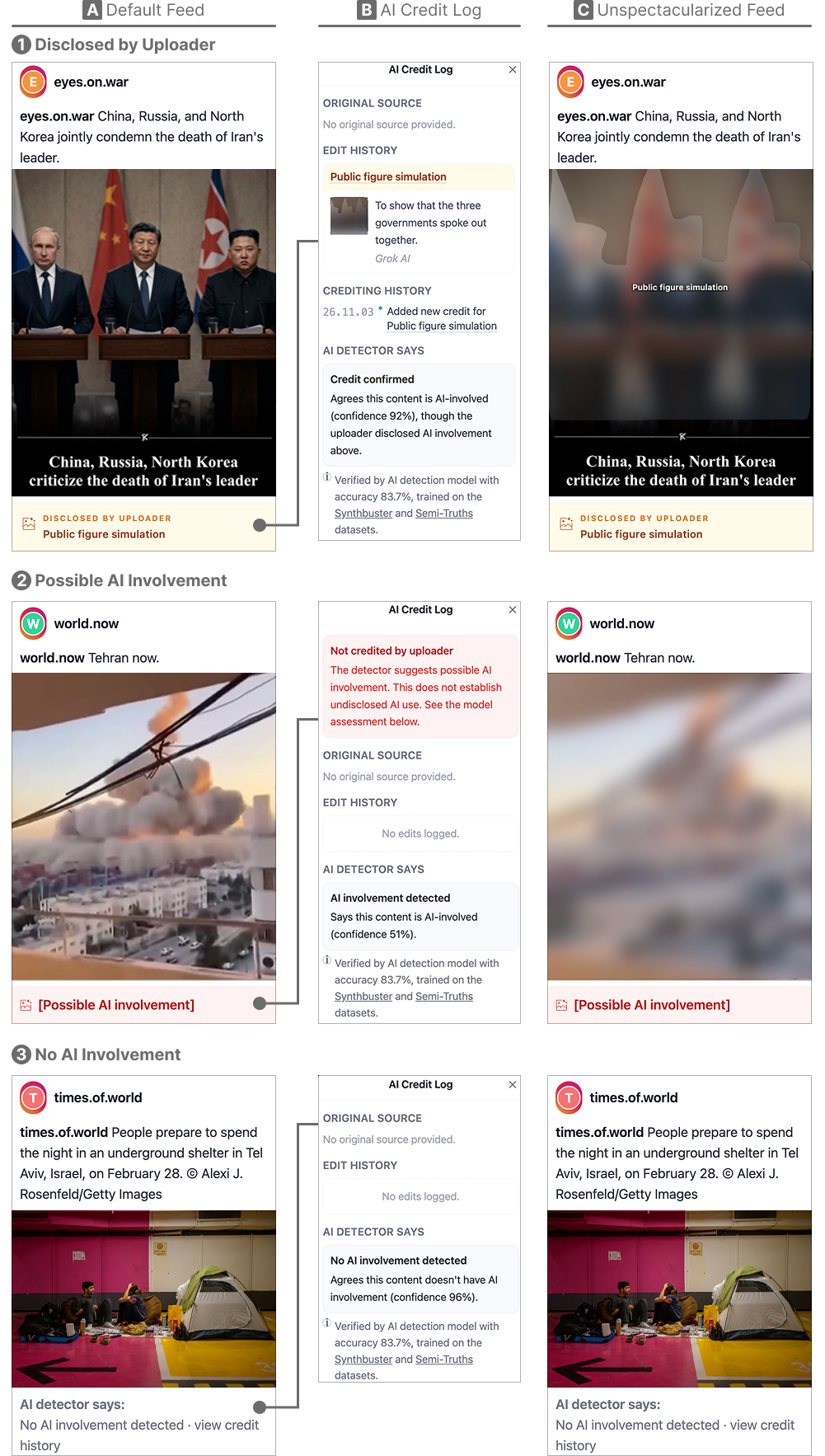}
        \caption{Unspectacular Feed Labeling System.}
        \label{fig:label}
    \end{minipage}
\end{figure*}


Based on these design goals, we present \textit{Unspectacular Feed} (Fig \ref{fig:interface}), connecting AI-use disclosure, automated detection, and creator reputation to encourage accurate reporting of AI involvement. 


\subsection{Interface Design}
\subsubsection{Disclosing AI}
After selecting an image, creators enable \textit{Is AI involved in this post?} to begin crediting. Because process-based labels communicate provenance without establishing veracity~\cite{feng2023provenance}, the interface captures both how and why AI was used (DG1, 2). Creators can optionally attach source images and generation chat logs, accommodating both edited photographs and fully generated images. These records support traceability and attribution, including reference performances reused in AI-generated entertainment~\cite{kyi2025governance}. Creators select AI techniques (Fig \ref{fig:interface} B), with descriptions adapted from Gemini, Grok, Suno, and Higgsfield documentation, and can specify each intervention's purpose, model, and prompt summary. For image-related techniques, a drawing tool marks altered regions (Fig \ref{fig:interface} C), which are blurred in Unspectacularized mode. Creators choose an overlay showing the technique, purpose, both, or custom text, keeping disclosure attached to the affected area.


\subsubsection{Viewing Credit History}
After publication, the baseline detector assesses the image. Posts display \textit{Disclosed by Uploader} with declared techniques, \textit{Possible AI Involvement} when detection flags undisclosed AI use, or \textit{No AI involvement detected} when neither source indicates AI use (Figure \ref{fig:label}). Clicking a label opens available sources, techniques and descriptions, marked regions, purposes, models, and prompts (Fig \ref{fig:interface} D). Following journalistic principles of acknowledging errors and explaining corrections~\cite{spj2014ethics}, the history preserves initial disclosures and subsequent additions, modifications, and deletions, making revisions inspectable as acts of accountability (DG3). The detection panel displays agreement with the disclosure, confidence, training data, and measured accuracy (DG2).


Agreement concerns AI presence; it does not verify individual edits or purposes. Unspectacularize lets viewers control exposure to AI-generated imagery while retaining post context (Fig \ref{fig:interface} D). 
Addressing viewers' filtering preferences and creators' concerns about dismissal, the toggle blurs AI-involved regions and overlays the technique, purpose, both, or custom text. Viewers inspect this context before choosing to reveal the regions (DG4).
Sharing within the platform or through external links defaults to the Unspectacularized version (Fig \ref{fig:interface} F). Although AI-origin labels can have limited effects on sharing intentions~\cite{wittenberg2025labeling,10.1145/3706598.3713171}, embedding context within images may help preserve it through reposting. The sharing preview includes the obscured image and credit history, allowing users to retain or omit disclosure context and consider their credibility as sharers (DG4).
The default feed also restricts capture of AI-involved posts and directs users to the Unspectacularized version (Fig \ref{fig:interface} E). This mechanism is intended to reduce circulation without disclosure context, although it cannot prevent subsequent removal or circumvention.


\subsubsection{Profile Disclosure Score}
Each profile summarizes the percentages of AI-involved and uploader-disclosed posts (Fig \ref{fig:interface} G). Orange markers indicate disclosed AI use; red markers indicate detector-flagged posts without disclosure. Signaling theory suggests that sustained, inspectable behavior can support credibility when it is difficult to imitate~\cite{donath2007signals}. Accumulated disclosure histories may therefore signal helping viewers decide whom to follow and are intended to give creators reputational incentives for consistent, accurate disclosure (DG4).

\subsection{AI-Image Detection Model}
We used the pretrained Haywoodsloan AI-image detector\cite{haywoodsloanDetector}, based on a SwinV2 vision transformer\cite{liu2022swinv2}, to identify potentially AIGC and complement creator-provided disclosures. We selected this baseline following preliminary comparisons that highlighted false-positive concerns. For evaluation, we used 400 images: 100 real and 100 generated images from OpenFake\cite{livernoche2025openfake}, and 100 matched real/AI-edited pairs from FragFake\cite{sun2025fragfake}. These datasets allowed us to assess both full-image generation and partial editing. The detector produces an AI-class score between 0 and 1, with scores of at least 0.5 classified as AI-generated, but does not localize the regions generated or altered by AI. This score reflects the model’s confidence. As shown in Table~\ref{tab:baseline-evaluation}, the baseline performed better on fully generated images than on partially edited images, suggesting localized modifications remain difficult to detect. Differences in dataset sources and image formats may also contribute to this gap. False positives remained present, and evaluation was limited to a modest, structured sample containing matched original/edit pairs. For the user study, we selected only AI-generated images flagged by the detector. This selection supported examining responses to detection disclosures but limits generalization to missed cases. We treat the baseline as a fallible signal that complements creator disclosures. 

As shown in Table~\ref{tab:baseline-evaluation}, the baseline performed better on fully generated images than on partially edited images, suggesting that localized modifications remain difficult to detect. However, differences in dataset sources and image formats may also contribute to this gap. False positives remained present, and evaluation was limited to a modest, structured sample containing matched original/edit pairs. For the user study, we selected only AI-generated images flagged by the detector. This selection supported examining responses to detection disclosures but limits generalization to missed or ambiguous cases. We therefore treat the baseline as a fallible signal that complements creator disclosures. 

\begin{table}[]
\centering
\caption{Haywoodsloan baseline evaluation at a fixed AI-class
threshold of 0.5. AI-generated or AI-edited content is the
positive class. All metrics except AUROC are percentages.}
\label{tab:baseline-evaluation}
\small
\begin{tabular}{lrrrrrrrr}
\toprule
Dataset & $N$ & Accuracy & AUROC & Precision & Recall & F1 & FPR \\
\midrule
OpenFake & 200 & 85.00 & 0.946 & 96.1 & 73.0 & 83.0 & 3.0 \\
FragFake & 200 & 64.50 & 0.758 & 85.4 & 35.0 & 49.6 & 6.0 \\
\textbf{Overall} & \textbf{400} & \textbf{74.75} &
\textbf{0.869} & \textbf{92.3} & \textbf{54.0} &
\textbf{68.1} & \textbf{4.5} \\
\bottomrule
\end{tabular}

\smallskip
\begin{minipage}{\linewidth}
    \footnotesize
    \raggedright
    Each dataset contains 100 real and 100 AI-positive images. OpenFake evaluates generated images; FragFake evaluates 100 matched original/edit pairs. FPR denotes the proportion of real images incorrectly classified as AI-generated. The evaluation contains 400 images across 300 dataset-defined groups; the pretrained model was not fine-tuned on this sample.
\end{minipage}
\end{table}

%% file: V3/06_User_Study.tex
\section{User Study}
We conducted two complementary studies with creators and viewers (N=15 each) to examine how \textit{Unspectacular Feed} supports AI-use disclosure and credibility assessment in conflict-related social media posts. Iran-US war-related images were selected to reflect real-world cases where photorealistic AI-generated media was actively circulating and misleading the audience. In this study, we aim to answer the following research questions: \textbf{RQ1.} How do creators articulate the techniques and purposes of AI use, and how does \textit{Unspectacular Feed} support or hinder clear and accurate crediting of their production processes? \textbf{RQ2.} How do viewers use credit histories and Unspectacularized mode to understand AI involvement and assess the trustworthiness of posts and uploaders? \textbf{RQ3.} How do creators and viewers interpret automated detection results, and how do these interpretations inform their trust in the detector and \textit{Unspectacular Feed}?

\subsection{Study Design}
\subsubsection{Participants}
We recruited 30 participants through Prolific: 15 creators and 15 viewers. Eligibility required participants to be at least 18, fluent in English, US residents, and active social media users. Creator screening additionally targeted interests in influencing, blogging, video creation, art, photography, design, or AI-assisted creative work, or employment as a content creator, journalist, or visual artist. Participants received USD \$15 for approximately 30 minutes. 

Among 14 creator pre-survey respondents, 9 identified as digital artists, 8 as issue-focused creators, and 4 as journalists (multiple selections permitted). All reported using generative AI, including 11 for text and seven for still images; 8 used it in approximately half or all of their content. Among 15 viewers, 13 encountered news or informational content at least daily, and 14 had encountered suspected AI-generated posts. All 14 used visual artifacts as recognition cues, while nine also relied on platform labels. Both groups reported high AI literacy, yet recognition varied across three AI-generated examples. 13 creators and 14 viewers identified the image with conspicuous artifacts as AI-generated, compared with 7 and 8 for the photorealistic person image, and 4 and 5 for the satellite image. Detailed participants screening criteria, characteristics, AI-literacy scores, and recognition measures appear in Appendix~\ref{app:user-study-participant}.

\subsubsection{Task}
We used hypothetical scenarios to situate disclosure and credibility judgments within a shared context, following prior HCI research~\cite{vaccaro2020contesting}.
\textbf{Creators} were guided to be in the role of an Instagram magazine editor-in-chief preparing coverage of an Iran--US conflict based on three reporter's email. Each participant received one fully AI-generated image and two AI-enhanced images, accompanied by reporter-provided descriptions of the purpose of AI use, original source material, and editing technique. Particiapnts first explored \textit{Unspectacular Feed} with three example posts. Then, creators uploaded three posts. Within the disclosure form, selecting an AI technique was required, while describing the purpose of AI use, uploading source material, and identifying edited regions were optional. After uploading, baseline detection system review the post and show the credit history. 
\textbf{Viewers} were guided to explore the feed and all provided contents are from creators they are following, containing nine posts about an Iran--US conflict: three with fully AI-generated images, three with partially AI-enhanced images, and three with original images. Among the fully AI-generated posts, one carried a detector-generated label and two carried uploader-provided disclosures. Among the partially AI generated posts, one carried a detector-generated label and two carried uploader-provided disclosures (Appendix \ref{app:study-tasks}). Participants explored the feed for three minutes using a checklist that prompted them to open the credit history, explore the unspectacularized feed, share content, view a creator profile, and capture a post. After this period, participants could proceed to a post-by-post survey assessing perceived authenticity, comfort with disclosure and system verification, and perceived accuracy of AI-region localization. 
After both creator viewer task, they were guided to complete a post-survey.

\subsubsection{Measures}
We analyzed open-ended responses and creators' disclosed purposes and prompt summaries using thematic analysis. Questionnaire responses and recorded interactions were analyzed separately for creators and viewers.
For creators, questionnaires assessed perceived usability, clarity of crediting fields, adequacy of representing different degrees of AI involvement, effort, disclosure concerns, and adoption intention. We also examined submitted disclosure records and classified saved polygon coverage as near-whole-image (at least 90\% of image area), partial-image (a nonempty selection covering less than 90\%), or no drawing. For viewers, questionnaires assessed perceived usability, content credibility, and sharing intention for selected fully AI-generated, partially AI-generated, and non-AI images, alongside perceived transparency, usefulness of regional and profile information, and adoption intention. Detailed questionnaire items appear in Appendix~\ref{app:user-study-post}.
We summarized individual ordinal ratings using response distributions, medians, and interquartile ranges. Within-participant comparisons across three or more conditions used permutation-based Friedman tests \cite{friedman1937ranks}. Follow-up paired comparisons used Wilcoxon signed-rank tests \cite{wilcoxon1945ranking}.
Creator logs were used to characterize submitted credit information, source-material inclusion, and elapsed upload duration. Viewer logs were used to characterize credit-history coverage, opening frequency, recorded exposure duration, post exposure, and sharing actions. Browsing-phase interactions were analyzed separately from survey-phase interactions. 

\subsection{Results}
We report how creators translated production information into disclosure (RQ1), how viewers assessed the credibility of posts and creators in relation to disclosure (RQ2), and how they responded to automated detection (RQ3). We also examine the effort involved in documenting and inspecting AI use, and how the system supported reflection on AIGC.

\subsubsection{\textbf{Creators varied in mapping and classifying AI generated Content}}
\label{sec:sec621}
We examined what the 15 creators' disclosure records across 45 submissions: the selected AI technique, explanatory text, supporting sources, marked regions, and the information selected for regional overlays. Creators could explain a production rationale, supply process evidence, or identify a region, but these elements did not always communicate the same account of how the image was made.
Importantly, creators interpreted \textit{used AI techniques} differently for the same post (see Table \ref{tab:region-disclosure}). C11 paired a simulated-generation prompt with enhancement framing, while C14 described enhancement as appropriate in the questionnaire but saved scene simulation. One creator selected audio cleanup for all three image-only posts. These inconsistencies suggest that descriptions of AI techniques should be precise and construct a societal norm around them, since personal interpretation of the technique is different. Especially, it shows the difficulty of compressing specific changes into a broad label. It also implies that creators can check for incorrect AI techniques; if there is no verification of the selected AI technique, the disclosure system can add wrong information. We are discussing establishing norms for AI techniques in the further Discussion section (\ref{sec:sec72})

We also found that participants can produce misinformation by adding wrong sources. For the radar task, which supplied neither an original image nor a chat link, 9 creators attached images and 3 submitted chat URLs; all 3 URLs referenced other tasks. These task-level differences distinguish retaining available evidence, omitting it, and supplying material whose relevance requires verification. Outside of attaching original sources, 21 submissions named a model in the model field, and 21 contained substantive prompt or instruction text under the preliminary coding rubric. These features did not necessarily coincide: C1 supplied source evidence while leaving model and prompt fields blank, whereas C3 supplied substantial prompt descriptions without source attachments.

The spatial scope of disclosing the AI generated region varied across participants. Table \ref{tab:region-disclosure} shows that even the fully generated image was not uniformly marked across its full extent, while some enhancement submissions marked nearly the entire image. These patterns show that creators make different decisions on how to describe AI generated regions. For the overlayed text on the disclosed region, 33 out of 45 submission were selected to overlay purpose, 8 technique + purpose, 2 technique only, and 1 custom text, showing creators valued disclosing purpose compared to technique.

\begin{table*}[]
\centering
\small
\renewcommand{\arraystretch}{0.7}
\caption{Creator disclosures by task ($N=15$ each). Counts for supporting materials may overlap.}
\label{tab:region-disclosure}
\begin{tabularx}{\textwidth}{
    @{}
    >{\raggedright\arraybackslash}p{2.6cm}
    >{\raggedright\arraybackslash}p{2.1cm}
    >{\raggedright\arraybackslash}X
    @{}
}
\toprule
Task & Disclosure aspect & Responses ($n$) \\
\midrule

Fully generated airstrike
& Region markings
& Near-whole image: 6; partial image: 3; no drawing: 6. \\
& AI techniques
& Scene simulation: 9; scene extension: 2; visual enhancement: 2; public-figure simulation: 1; audio cleanup: 1. \\

\midrule
Provenance-uncertain radar image
& Region markings
& Near-whole image: 4; partial image: 8; no drawing: 3. \\
& AI techniques
& Scene simulation: 8; scene reconstruction: 4; color recalibration, visual stabilization, and audio cleanup: 1 each. \\

\midrule
Reporter-described gravesite enhancement
& Region markings
& Near-whole image: 4; partial image: 4; no drawing: 7. \\
& AI techniques
& Color recalibration: 6; visual enhancement: 3; audio cleanup: 2; object removal, visual stabilization, scene simulation, and speaker identification: 1 each. \\

\bottomrule
\end{tabularx}

\smallskip
\begin{minipage}{\textwidth}
\footnotesize
\end{minipage}
\end{table*}

On the other hand, creators used purpose fields to explain production decisions and qualify uncertain evidence. Thematic coding identified 14/45 fields containing a rationale, 16 combining an operation and a rationale, and four naming only an operation or technique. C1's \textit{To give a glimpse of what could be} expressed an illustrative purpose, whereas C9's \textit{removed excavators} identified an operation but omitted the color change requested in the accompanying prompt. All six provenance caveats occurred in the radar task. C1 wrote, \textit{Please note the original image is believed to be an AI image since no image of the attack has been documents.} This qualified the image's evidentiary status without claiming verified knowledge of its production. These responses show that the purpose field elicited rationales, descriptions of alterations, and warnings about uncertain provenance, providing context beyond a technique label alone.

\subsubsection{\textbf{Disclosure credibility did not imply trust in the depicted content}}

To answer RQ2, we compared viewers' in-task authenticity-trust ratings with their comfort about each post's disclosure. We also examined post-survey credibility and sharing ratings for fully generated, partially generated, and non-AI examples. The motivating question was whether credibility tracked the kind of AI involvement, or whether a well-received disclosure could make even a fully generated depiction credible. 
Across 135 paired post evaluations, 26 combined high disclosure comfort (scores 4--5) with low authenticity trust (scores 1--2); 10 of 15 viewers exhibited this combination at least once. For the creator-declared world-leader image (Appendix \ref{app:study-tasks} Post 2), mean disclosure comfort was 3.53/5 while mean authenticity trust was 1.73. For the creator-declared hypothetical airstrike (Appendix \ref{app:study-tasks} Post 4), the corresponding means were 3.40 and 2.47. The creator-declared gravesite image (\ref{app:study-tasks} Post 6) received higher trust (3.47) alongside comfort of 3.87. These within-post contrasts show that viewers could be comfortable with a disclosure while remaining skeptical of the depiction.

Likewise, the post-survey examples differed in credibility. Mean credibility was 1.53/5 for the fully generated example, 2.33 for the partially generated example, and 4.27 for the non-AI example; mean sharing likelihood was 1.33, 1.73, and 3.40, respectively. All 15 viewers rated the non-AI example more credible than the fully generated example. However, image content and presentation varied alongside AI involvement and disclosure. 

Participants' comments explain why credibility of disclosure doesn't connect to credibility of image. V3 said that learning about AI involvement in the world-leader image \textit{made me question how real the image was.} V6 instead said \textit{the credits helped to solidify those opinions}, suggesting confirmation. V13 stated, \textit{If it's AI generated, I inherently trust it less, even if it's proactively disclosed by the publisher.} These accounts suggest that disclosure could revise or confirm a judgment without overcoming broader skepticism. Lower authenticity trust after understanding how an image was made may be an appropriate consequence of disclosure, rather than a failure to make the post credible. Also, creators' profile information about their disclosure history and AI-involved post counts were a related but separate judgment: 11 viewers agreed that profile information helped assess creator reputation, and 14 said it influenced trust of their post. To answer RQ2, viewers distinguished the credibility of a depiction from their response to its disclosure and its creator; an acceptable disclosure did not consistently translate to AI-generated content's credibility. 

\subsubsection{\textbf{Automated detection informed judgments, but continued skepticism}}

For RQ3, we examined how viewers evaluated detectors' verification and described uncertainty about automated decisions. 14 of 15 reported understanding the distinction between creator declarations and automated flags, and 12 believed the credit/detection system could accurately identify AI content. Nevertheless, favorable perceptions did not imply unqualified acceptance. The automatically flagged radar-strike post received mean authenticity trust of 1.80/5 and disclosure comfort of 3.27; the automatically flagged Tehran post (Appendix \ref{app:study-tasks} Post 1) received 2.07 and 3.13. These ratings concern the combined post and disclosure context, not an isolated test of detector confidence or correctness.

V5's comments expose the unresolved basis of some judgments. On one post, this viewer wrote, \textit{I don't know how Synthbuster and Semi-Truths work. It does appear very confident though.} On another, they noted, \textit{They seem pretty confident that it was not machine generated.}. The numerical estimate was salient despite uncertainty about the methods. This could indicate skepticism about an authoritative presentation. V10 raised the possibility of a false AI flag directly: \textit{What if AI is wrong? If there is actually real content being mistaken as AI?} These cases suggest that automated detection added information for evaluation but did not settle whether an image or declaration should be believed.

\subsubsection{\textbf{Creators anticipated credibility benefits from disclosure effort}}

Creators' responses separated the burden of disclosure from its anticipated value. 7 of 15 considered crediting time-consuming, while 11 believed that the required effort added credibility. 5 endorsed both statements, whereas 4 endorsed the credibility benefit while disagreeing that the process was time-consuming. 12 considered the requirement proportionate, but 8 said detailed disclosure could discourage future AI use. These responses suggest that a disclosure system could change how creators are using AI in their production workflow. C5 articulated an evidence-oriented account of the benefit: showing the original image and explaining \textit{the process and purpose of using AI to alter that image} could \textit{build trust with your audience.}.

The usability evaluation helps locate the burden within the creator workflow. 12 of 15 creators agreed that the tool was easy to use, but 5 found it cumbersome and 6 anticipated needing technical support. Their ten-item System Usability Scale\cite{brooke1996sus} scores averaged 59.67 ($SD = 22.00$). Median in-app completion time was 17.15 minutes, including navigation, waiting, and pauses. C11 reported, ``\textit{There was nothing confusing or difficult. It was very user friendly.}''  9 creators would use the system in their workflow and ten would recommend it. Disclosure also carried potential privacy or competitive costs. 13 creators felt comfortable sharing prompts and conversation logs in the task, while 7 expressed concern about prompt reuse. 

Also, from the viewer side, 13 viewers found the system easy to use, and 12 disagreed that viewing credits took too much time. 12 would use a similar system on their own platforms. V5's appreciation that a creator ``at least admitted it and explained how they edited the image'' suggests acknowledgment of candor. These findings suggest that creators anticipated credibility benefits from actively disclosing their AI use.

\subsubsection{\textbf{Disclosure supported reflection on interpreting AI generated media}}
Participants commonly reported that the system helped them articulate or interpret AI use. 13 creators agreed that seeing the viewer-facing presentation helped them describe their own AI involvement. 12 viewers reported understanding how AI had been used, 14 said the information helped them decide how much to trust an image, and 14 said it helped them decide whether to engage.

Participants' comments identify a more specific reflective process. C1 found that predefined options make them to reflect what they were using for actual content production. From the viewer side, V11 described attention to subtle alterations: \textit{When things are not pointed out to you, you tend to not think about the little things.}, and V3 described \textit{The grave photo was surprising. I had no idea if real images can be alter detail with AI}. V14 said \textit{The image of the graves made me trust the post a bit less, because the AI-editing seemed inauthentic and unnecessary.}, showing that disclosures also informing viewers to evaluate if certain processes were necessary. These accounts suggest that the system created new signals to re-consider the production decisions that might otherwise remain unnoticed from the viewer side.

%% file: V2/07_Discussion.tex
\section{Discussion}

\subsection{Establishing Shared Norms for AI Disclosure}
Our findings suggest that disclosure requires shared understandings of what counts as AI involvement and why it matters. Creators valued options that distinguished production practices: C7 explained, \textit{it was helpful that there were pre generated options to choose from. It gave me a simplified way of telling how the AI was used.} Yet these categories remained open to interpretation. In the formative study, P6 regarded AI-assisted object removal as routine editing because similar results can be achieved manually in Photoshop. This raises a question about disclosure thresholds: \textit{should they depend on the tool used, the extent of automation, or the changes made to the depicted scene?}

Policies draw these boundaries differently. The EU AI Act exempts certain assistive edits from provider marking and adapts disclosure for artistic or satirical deepfakes~\cite{euaiact2024}; YouTube exempts unrealistic imagery and minor aesthetic edits~\cite{youtubeAIdisclosure}; and Adobe Stock distinguishes primary-subject alterations from some background extensions and object removals~\cite{adobeStockGenAI}. Only 9 of 40 platforms explicitly defined AI-generated content or related concepts~\cite{gao2026governance}. Alongside evidence that anticipated judgment can encourage concealment~\cite{zhang2025secret}, these differences motivate clearer criteria that creators can apply across production workflows.


Disclosure also leaves judgments of acceptable use unresolved. Process- and harm-based labels address complementary questions~\cite{wittenberg2025labeling}, as C3 illustrated: \textit{Pushing fake images of war should not be made and pushed onto the public regardless if it's labeled or not.} News publishers' adoption of entertainment-oriented social media formats further complicates distinctions between illustration, satire, and evidentiary imagery~\cite{newman2022tiktok}. International journalistic principles of verification and correction~\cite{ifj2019charter} offer a foundation for examining how AI disclosure standards could support cross-border interpretation while accommodating cultural differences. These unresolved boundaries also motivate \textit{personalized Unspectacularization} as a future research direction. Alongside shared disclosure standards, viewers could configure blurring according to the technique, extent of AI involvement, or content context. For example, preferences could distinguish AI-generated memes, AI-assisted advertisements, and photorealistic scenes, with contextual overlays available before viewers choose to reveal the affected regions. Where societal norms for AI disclosure remain unsettled, these controls could accommodate individual exposure preferences while maintaining platform-wide disclosure requirements.

\subsection{Communicating Detector Uncertainty}
Our findings highlight how detector errors can become reputational harms. Photojournalists have lost employment or had their archives withdrawn following alterations such as removing seemingly incidental objects~\cite{apPhotoAlteration}, darkening explosion smoke~\cite{bdcAdnanHajj}, or combining photographs taken at different moments~\cite{latimes2003walski}. Detectors that attribute manipulation should likewise make their assessments reviewable and correctable. A false AI flag can cast doubt on legitimate work and creator. It can also burden creators with demonstrating human authorship or providing production records to rebut automated suspicion. Our current prototype does not adequately protect against or repair these harms. Text-detection research illustrates potentially unequal consequences: Liang et al. found that tested detectors disproportionately misclassified non-native English writing~\cite{liang2023bias}, while Chambers and Kelley identified higher flagging rates in a ``likely-autistic'' Reddit corpus using a GPT-2 detector~\cite{chambers2025autistic}. These findings highlights how we communicate the consequences of detection errors. Independent detection nevertheless remains useful when production information is withheld or disputed, including strategic denials of authentic evidence through the liar's dividend~\cite{chesney2019deepfakes}. However, incorrect predictions can undermine human judgment~\cite{groh2022crowds}. V5's attention to confidence scores alongside doubts about the methods illustrates why detection must remain contestable.

Therefore, uncertainty communication should identify a flag's basis, evaluation conditions, and failure modes. Confidence displays, deliberation prompts, and collaborative annotation offer complementary support, with limits to accuracy and usability~\cite{zhang2020calibration,bucinca2021trust,10.1145/3772318.3790501}. Future work should combine these approaches with detectors that adapt to emerging generators~\cite{li2023cddb}, preserving assessment histories, corrections, and reputation review. Evaluation should examine false accusations, reputational recovery, and the burden of interpreting changing assessments.

\subsection{Potential Harms of Voluntary Disclosure}
\label{sec:sec72}
Voluntary disclosure introduces uncertainty because creators' accounts may be mistaken, incomplete, or strategically framed. As reported in Section~\ref{sec:sec621}, one creator selected \textit{audio cleanup} for all three image-only posts, while other records contained inconsistent techniques, prompts, and sources. They do not establish intentional deception, but show that completed disclosures can create an appearance of transparency without a coherent production account.

AI labels can function as conventional signals whose reliability depends on verification and social enforcement~\cite{donath2007signals}. Handicap signaling suggests that costs can sustain reliability when deceptive signaling becomes sufficiently disadvantageous~\cite{grafen1990handicaps}. Perceived effort could nevertheless be weaponized: an uploader might disclose minor object removal from a fully generated image while concealing its simulated origin, gaining credibility through selective transparency. Conversely, anticipated judgment and trust penalties may discourage accurate disclosure~\cite{zhang2025secret,schilke2025transparency}. Rewarding disclosure effort, completeness, or frequency alone could therefore favor persuasive self-presentation over reliable evidence.

Future systems should distinguish creator statements, authenticated provenance, detector assessments, and independently reviewed corrections. Studies should examine accidental errors and deliberate misrepresentation separately, assessing their effects on viewer judgments and creator reputation. They should also investigate how verification and accountability can constrain deceptive signaling without demanding excessive private production data or treating completeness as evidence of honesty.

%% file: V1/08_Limitation.tex
\section{Limitations}
We used images related to the US--Iran conflict to examine AI-generated media in a high-stakes context. Participants' existing beliefs and attitudes toward the conflict may have shaped their interpretations and trust judgments. Additionally, all participants were based in the US. Their perspectives may reflect culturally situated expectations about AI-generated media and AI use. Studies across cultural and geographic contexts would help examine the transferability of our findings.

%% file: V1/09_Conclusion.tex
\section{Conclusion}
This paper presents \textit{Unspectacular Feed}, an interactive system connecting process-based crediting, automated detection, and creator disclosure histories to support AI labels as contestable credibility signals. Creators document production purposes, sources, models, and AI-altered regions; viewers inspect these records with detection results and contextual overlays that remain available through sharing and profiles. In two studies with creators and viewers ($N=15$ each), creators valued communicating production rationales, although disclosure records varied in consistency. Viewers used disclosures to assess trustworthiness and reported greater awareness of partial AI editing, while remaining skeptical of depicted content and detector reliability. These findings suggest that contextual disclosure can support informed scrutiny of AI-mediated media. Future work should examine how to design system based on differing norms of AI involvement, communication of detection uncertainty, and potential misuse of voluntary disclosure.

%% file: V2/10_Gen_AI.tex
\section{Ethics and Privacy Statement}
This study was exempt from institutional review board (IRB) approval. The user study was conducted anonymously through Prolific, and we did not collect personally identifiable information. Study images were either generated by the authors or sourced online that was identified by journalists as AI-generated or AI-altered.

Because the materials depicted conflict-related scenes, participants in both formative study and user study were informed that they could discontinue participation if they felt uncomfortable. Content warnings appeared before the study and again before tasks involving exposure to AI-generated or partially AI-generated conflict imagery.

\section{Generative AI Disclosure}
Generative AI was not used to substantially shape the study’s methodology, analysis, or conclusions. ChatGPT assisted with coding the user-study website, improving grammar, and formatting tables.

%% file: V2/10_Appendix.tex
\clearpage
\onecolumn
\appendix

\section{Formative Study Materials}

\label{app:formative-participants}
\subsection{Formative Study Participants}

\begin{table}[H]
\centering
\small
\renewcommand{\arraystretch}{1.15}
\setlength{\tabcolsep}{6pt}

\begin{tabularx}{\linewidth}{
    @{}l
    >{\raggedright\arraybackslash}X
    >{\raggedright\arraybackslash}X@{}
}
\toprule
\multicolumn{3}{@{}l}{\textbf{A. Participant characteristics}} \\
\addlinespace[4pt]
Participant & Type & Age range \\
\midrule
P1 & Viewer & 30--39 \\
P2 & Viewer & 30--39 \\
P3 & Creator & 30--39 \\
P4 & Viewer & 30--39 \\
P5 & Viewer & 30--39 \\
P6 & Creator & 19--29 \\
\bottomrule
\end{tabularx}

\medskip

\begin{tabularx}{\linewidth}{
    @{}l
    >{\raggedright\arraybackslash}X
    >{\raggedright\arraybackslash}X@{}
}
\toprule
\multicolumn{3}{@{}l}{\textbf{B. Creator backgrounds}} \\
\addlinespace[4pt]
Participant & Content produced & Generative AI uses \\
\midrule
P3 & New media art. & Planning, drafting, and prototyping. \\
\addlinespace[3pt]
P6 & Finance and economics coverage for a
broadcaster-affiliated Instagram magazine.
& Still-image generation and content planning. \\
\bottomrule
\end{tabularx}

\smallskip
\begin{minipage}{\linewidth}
\footnotesize
Both creators published on Instagram and reported
occasional generative AI use.
\end{minipage}

\medskip

\begin{tabularx}{\linewidth}{
    @{}l
    >{\raggedright\arraybackslash}X
    >{\raggedright\arraybackslash}p{0.27\linewidth}
    >{\raggedright\arraybackslash}p{0.15\linewidth}@{}
}
\toprule
\multicolumn{4}{@{}l}{\textbf{C. Viewer backgrounds}} \\
\addlinespace[4pt]
Participant & Content consumed & Platforms
& News exposure \\
\midrule
P1 & News, art, lifestyle, and entertainment.
& YouTube, Instagram, Facebook & Daily \\
\addlinespace[3pt]
P2 & News, politics, social issues, and education.
& YouTube, Facebook, LinkedIn & Daily \\
\addlinespace[3pt]
P4 & Art, lifestyle, and entertainment.
& Instagram & About weekly \\
\addlinespace[3pt]
P5 & Industry and research news.
& LinkedIn & Daily \\
\bottomrule
\end{tabularx}

\end{table}

\subsection{Disclosure Activity}
\label{app:formative-disclosure}
\begin{figure}[H]
    \centering
    \includegraphics[width=\linewidth]{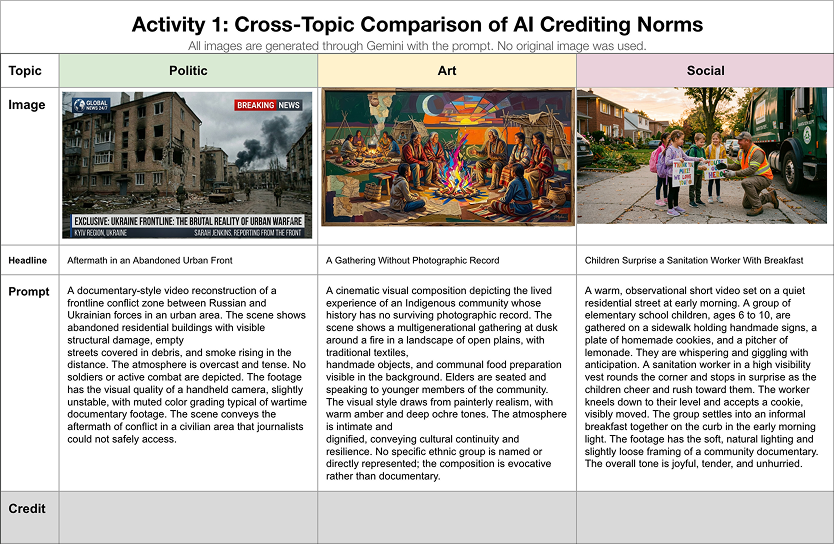}
    \caption{Activity 2-1: Participants assumed the role of an Instagram magazine editor-in-chief receiving three reports with AI-generated images, reasons for AI use, and generation prompts. They drafted disclosures using this information. They were guided like this: \textit{For this activity, imagine that you are the lead editor of an Instagram based magazine, and that someone has submitted these images for the news section. The first is an AI generated war image that simulates events in a conflict zone that reporters could not safely access. The second is an art piece that incorporates an AI generated scene to depict an unwritten history that no photographic record could capture. The third depicts a local neighborhood event, using an AI generated image to protect the portrait rights of the children involved. As the editor responsible for publishing these pieces, I am interested in how you would credit the AI involvement in each. Imagine you are the one uploading them. Could you write a comment for each that credits the AI involvement as you would in practice?}}
    \label{fig:appx-formative-activity-2-1}
\end{figure}

\begin{figure}[H]
    \centering
    \includegraphics[width=\linewidth]{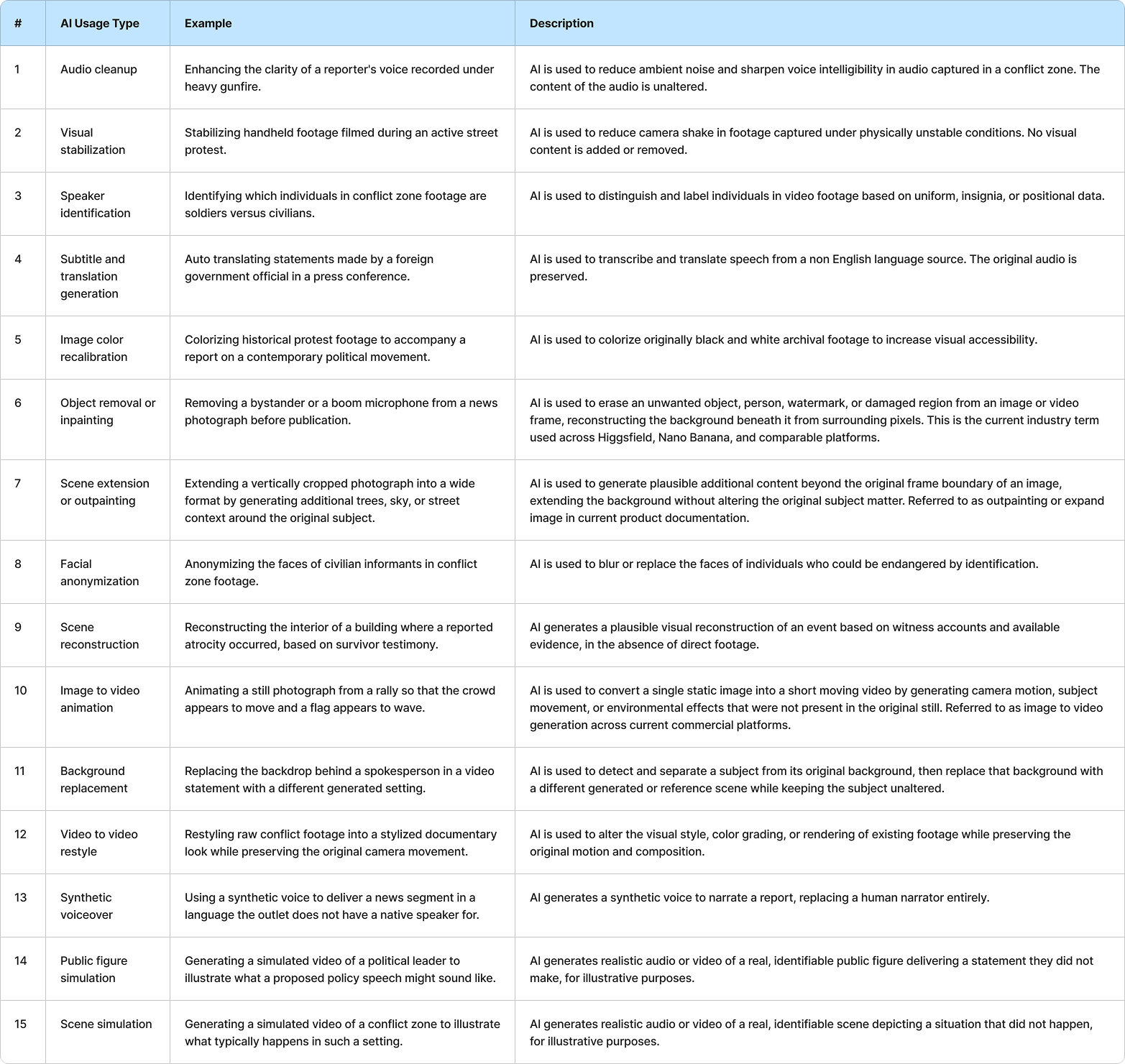}
    \caption{Activity 2-2: Participants evaluated 15 AI techniques and discussed how their use should be disclosed. Techniques and descriptions were adapted from Gemini, Grok, Suno, and Higgsfield documentation. They were guided like this: \textit{We will now move on to a broader set of examples. [Showing the taxonomy] Here are 11 content examples. All of them cover the issue of reporting the conflict zone and each indicates where AI has been used. Imagine you are scrolling through your feed and you encounter each of these. Could you sort them into three groups: (1) content you would want to see in your feed, (2) content you would not want to see, and (3) content you are uncertain about?}}
    \label{fig:appx-formative-activity-2-2}
\end{figure}

\section{User Study Materials}
\subsection{Participants Information}
\label{app:user-study-participant}

\begin{table}[H]
\centering
\small
\caption{Creators' production backgrounds and viewers'
consumption backgrounds. Listed categories reproduce
those reported in the pre-survey summary.}
\label{tab:user-content-background}
\begin{minipage}[t]{0.48\textwidth}
\vspace{0pt}
\centering
\textbf{Creators ($n=14$)}\par\smallskip
\begin{tabularx}{\linewidth}{
@{}>{\raggedright\arraybackslash}Xr@{}}
\toprule
Characteristic & $n$ \\
\midrule
\multicolumn{2}{@{}l}{\textit{\textbf{Creator roles}}} \\
Digital artist & 9 \\
Issue-focused creator & 8 \\
Journalist & 4 \\
\addlinespace
\multicolumn{2}{@{}l}{\textit{\textbf{Publishing platforms}}} \\
YouTube & 10 \\
TikTok & 8 \\
Facebook & 7 \\
Instagram & 7 \\
\addlinespace
\multicolumn{2}{@{}l}{\textit{\textbf{Content revenue}}} \\
Earned revenue & 6 \\
\addlinespace
\multicolumn{2}{@{}l}{\textit{\textbf{Monthly revenue among earners ($n=6$)}}} \\
USD~50--99 & 1 \\
USD~500--999 & 3 \\
USD~1,000 or more & 2 \\
\bottomrule
\end{tabularx}
\end{minipage}\hfill
\begin{minipage}[t]{0.48\textwidth}
\vspace{0pt}
\centering
\textbf{Viewers ($n=15$)}\par\smallskip
\begin{tabularx}{\linewidth}{
@{}>{\raggedright\arraybackslash}Xr@{}}
\toprule
Characteristic & $n$ \\
\midrule
\multicolumn{2}{@{}l}{\textit{\textbf{Content consumed}}} \\
Political coverage & 13 \\
Personal updates & 13 \\
Social issues & 12 \\
Hobby-related content & 12 \\
\addlinespace
\multicolumn{2}{@{}l}{\textit{\textbf{Platforms used}}} \\
YouTube & 15 \\
Reddit & 14 \\
Instagram & 13 \\
TikTok & 9 \\
Facebook & 9 \\
\addlinespace
\multicolumn{2}{@{}l}{\textit{\textbf{News/information practices}}} \\
Encountered at least daily & 13 \\
Never shared & 4 \\
Shared several times weekly or more & 7 \\
\bottomrule
\end{tabularx}
\end{minipage}
\end{table}

\begin{table}[H]
\centering
\small
\caption{Creators' generative AI use and viewers' prior
exposure to suspected AI-generated content.}
\label{tab:user-ai-background}
\begin{minipage}[t]{0.48\textwidth}
\vspace{0pt}
\centering
\textbf{Creators ($n=14$)}\par\smallskip
\begin{tabularx}{\linewidth}{
@{}>{\raggedright\arraybackslash}Xr@{}}
\toprule
Characteristic & $n$ \\
\midrule
Used AI to create or edit content & 14 \\
Used AI in approximately half or all content & 8 \\
\addlinespace
\multicolumn{2}{@{}l}{\textit{AI uses}} \\
Body or caption text & 11 \\
Still images & 7 \\
\addlinespace
\multicolumn{2}{@{}l}{\textit{AI tools}} \\
ChatGPT & 10 \\
Gemini/Nano Banana & 9 \\
\bottomrule
\end{tabularx}
\end{minipage}\hfill
\begin{minipage}[t]{0.48\textwidth}
\vspace{0pt}
\centering
\textbf{Viewers ($n=15$)}\par\smallskip
\begin{tabularx}{\linewidth}{
@{}>{\raggedright\arraybackslash}Xr@{}}
\toprule
Characteristic & $n$ \\
\midrule
Encountered suspected AI content & 14 \\
Unsure about prior exposure & 1 \\
\addlinespace
\multicolumn{2}{@{}l}{\textit{\textbf{Formats encountered ($n=15$)}}} \\
Video & 13 \\
Still images & 13 \\
Body or caption text & 10 \\
Engagement text & 8 \\
Music or audio & 7 \\
\addlinespace
\multicolumn{2}{@{}l}{\textit{\textbf{Recognition cues ($n=14$)}}} \\
Visual artifacts or irregularities & 14 \\
Automated platform labels & 9 \\
\bottomrule
\end{tabularx}
\end{minipage}
\end{table}

\begin{table}[H]
\centering
\small
\caption{Self-reported AI literacy and identification of
three AI-generated images.}
\label{tab:user-literacy-recognition}
\begin{tabularx}{\textwidth}{
@{}>{\raggedright\arraybackslash}Xrr@{}}
\toprule
Measure & Creators ($n=14$) & Viewers ($n=15$) \\
\midrule
AI literacy: $M$ ($SD$), out of 5
& 4.21 (0.50) & 4.07 (0.63) \\
\addlinespace
\multicolumn{3}{@{}l}{\textit{Images identified as AI-generated}} \\
Conspicuous visual artifacts & 13/14 & 14/15 \\
Photorealistic person & 7/14 & 8/15 \\
Satellite image & 4/14 & 5/15 \\
Overall across three images & 24/42 & 27/45 \\
\bottomrule
\end{tabularx}

\smallskip
\begin{minipage}{\textwidth}
\end{minipage}
\end{table}

Political affiliation and attitudes toward US military
spending were collected to contextualize responses to
the conflict-related materials.

\begin{table}[H]
\centering
\small
\caption{Self-reported political attitudes.}
\label{tab:user-political-background}
\begin{tabularx}{\textwidth}{
@{}>{\raggedright\arraybackslash}Xrr@{}}
\toprule
Characteristic & Creators ($n=15$) & Viewers ($n=15$) \\
\midrule
\multicolumn{3}{@{}l}{\textit{\textbf{Political affiliation}}} \\
Democrat & 6 & 10 \\
Independent & 6 & 3 \\
Republican & 2 & 2 \\
Other affiliation & 1 & 0 \\
\addlinespace
\multicolumn{3}{@{}l}{\textit{\textbf{US military-spending preference}}} \\
Reduce & 7 & 12 \\
Maintain & 4 & 2 \\
Increase & 4 & 1 \\
\bottomrule
\end{tabularx}
\end{table}

\subsection{Pre-Survey}
\label{app:formative-procedure}
\subsubsection{Creator Questions}
\begin{enumerate}
    \item What is your Prolific ID?
    \item Which category best describes the content you produce? Please select all that apply.
    \item Please briefly specify the nature of your content. For example: film reviewer, member of a broadcast newsroom's digital team, video artist, or independent music creator.
    \item Which platforms do you use to publish your content? Select all that apply.
    \item How frequently do you upload the content in your social media account?
    \item Do you earn money from the content you post, such as through ads, sponsorships, subscriptions, or affiliate links?
    \item [Optional] About how much money do you earn per month from your social media content? If your income varies, please estimate your monthly average over the past three months.
    \item [Optional] If you feel comfortable, please share a link to your main public account or channel so the researcher can review your published content. Sharing this link may reveal your identity, but researchers will not use this information or present this data in a research findings. You may leave this question blank.
    \item How frequently do you use generative AI when creating or editing content, including generating entire pieces or altering parts of existing content?
    \item If you use generative AI in your process, for what purposes do you use it? Select all that apply.
    \item Which AI tools did you use to create the content you described above? Select all that apply.
    \item [Optional] If you do not use generative AI in your content creation process, please briefly describe your reasons.
    \item [Optional] Please submit the example of the AI generated content that you’ve uploded to the social media(screenshot of the post)
\end{enumerate}

\subsubsection{Viewer Questions}
\begin{enumerate}
    \item What is your Prolific ID?
    \item Which of the following types of content do you most often consume on social media? Select all that apply.
    \item Which of the following social media sites do you regularly use (at least once a week)? Please select all that apply.
    \item About how frequently do you encounter news, issue coverage, or informational content in your social media feeds?
    \item About how frequently do you share news, issue coverage, or informational content in your social media feeds?
    \item Have you previously encountered a social media post that you knew or suspected involved AI generated content?
    \item If yes, how was AI used in the content you encountered? Select all that apply.
    \item If yes, what signal led you to recognize that the content was AI generated? Select all that apply.
    \item [Optional] If yes, please briefly describe the content you encountered. What was it about, which parts did you believe were AI-generated or altered?
    \item [Optional] What was your reaction or response to that content?
    \item [Optional] Please submit the screenshot of AI generated content that you encountered in social media.
\end{enumerate}

\subsubsection{Shared Definition and Questions}

\textbf{Definition presented before AI-use/exposure questions.}
In this study, ``AI-generated content'' refers to posts containing images, videos, text, or audio created or modified, entirely or in part, using generative artificial intelligence (AI). People may guide the process through prompts, provide source material, edit the results, or select the final output.

This includes:
\begin{itemize}
    \item Fully AI-generated content: Generative AI creates the entire image, video, text, or audio, often in response to human instructions.
    \item Partially AI-generated content: Generative AI modifies existing material by adding, replacing, removing, or reconstructing parts of it.
\end{itemize}

\textbf{Attention check and political attitudes}
\begin{enumerate}
    \item To show that you are reading carefully, please select ``Disagree'' for this statement.
    \item Generally speaking, do you consider yourself a...
    \item I believe US government should invest more money in the military.
\end{enumerate}

\textbf{AI literacy}
\begin{enumerate}
    \item I feel confident identifying when an image or video has been generated or edited using AI.
    \item I understand how generative AI tools such as Midjourney, DALL E, or ChatGPT create images.
    \item I am aware of methods or tools that can be used to detect AI generated content.
    \item Before sharing images or videos, I regularly verify the origin of visual content.
\end{enumerate}

\textbf{Image recognition.}
The following questions were repeated for each of three images.
\begin{enumerate}
    \item Do you think this image was generated using AI?
    \item [Optional] What features of the image influenced your answer? Please describe what you noticed.
\end{enumerate}

\textbf{Authenticity disputes}
\begin{enumerate}
    \item How easy or difficult would it be for someone to convince others that a real photo or video was AI-generated?
\end{enumerate}

\subsection{Task}
\label{app:study-tasks}

\subsubsection{Creator Task}

\paragraph{Study guideline.}
\begin{quote}
\itshape
You are the Instagram magazine chief editor. This week, you are preparing
the special cover for a major current event. Three reporters from your
team submitted content. As chief editor, you have the authority to publish
content to Instagram. You will review three pieces of content, each
including a visual element that involves AI generated material. Each
reporter has noted the purpose of using AI, the technique used, and the
original image or prompt. Using this information, complete all three
uploads. Each post must disclose AI usage and include at least one AI
usage credit.
\end{quote}

\begin{table*}[t]
\centering
\caption{Creator task materials for the three uploads.}
\label{tab:creator-task-materials}
\small
\renewcommand{\arraystretch}{1.2}
\begin{tabularx}{\textwidth}{
  @{}p{0.20\textwidth}*{3}{>{\raggedright\arraybackslash}X}@{}
}
\toprule
\textbf{Material} & \textbf{Task 1} & \textbf{Task 2} & \textbf{Task 3} \\
\midrule
Source image
  & \includegraphics[width=\linewidth]{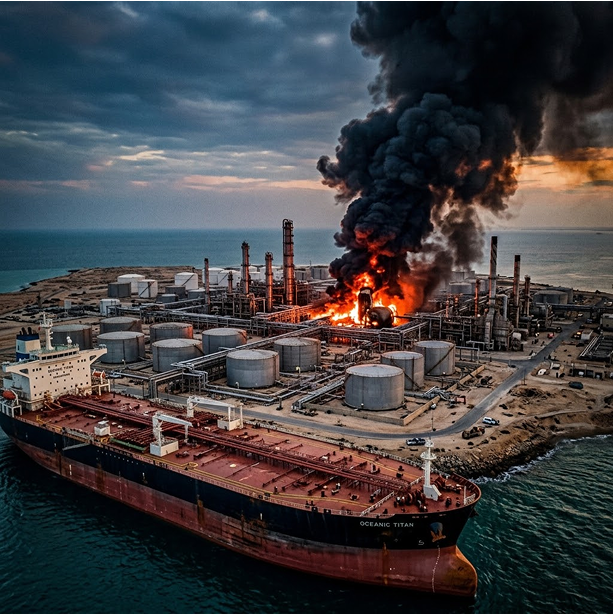}
  & \includegraphics[width=\linewidth]{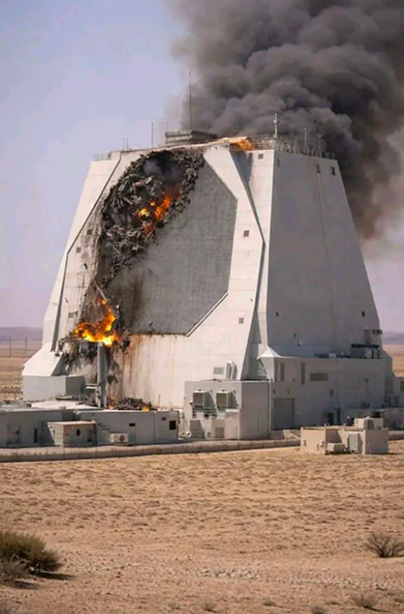}
  & \includegraphics[width=\linewidth]{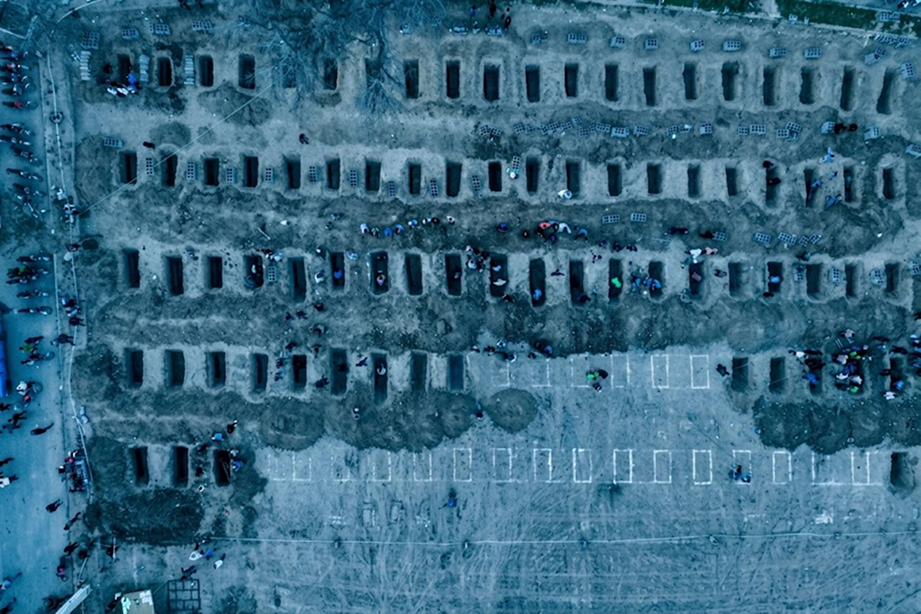} \\
Image type
  & Fully AI-generated & Partially AI-generated & Partially AI-generated \\
Reporter's email 
  & Hello. I'm submitting this week's cover, which addresses the US military's recent warnings about a potential airstrike on an Iranian island known for oil refining. The actual strike has not happened. I used an AI model to simulate what it would look like if the US military carried out the strike they have been warning about, in order to raise awareness of what this threat would actually look like if realized. I used Gemini's image generation model to fully generate this scene. Attached is the chat log documenting the generation process [link]. Let me know if you would like to discuss anything further. Thank you.
  & Hi, I am flagging a viral image currently circulating on social media that purports to show a strike on the U.S. AN/FPS-132 radar system in the Gulf. Iranian outlets, including the Tehran Times (citing the IRGC public relations office), are reporting that the 5,000-kilometer-range radar was destroyed. However, an initial visual assessment suggests the accompanying imagery circulating online is likely AI-generated rather than authentic footage. But I think since there is no photo for this attack, we can use it with the contents warning.
  & Hi. Following up on the reporting regarding the 175 Iranian elementary schoolgirls, I am submitting an aerial image of the victims' gravesite, originally sourced from AFP Photo / Iranian Press Center. Please note: Because the original drone photograph was low-resolution, I used Gemini Nanobanana[link] for color grading and visual enhancement to improve clarity. I have attached both the original, unaltered file and the enhanced version so the desk can review the adjustments against our photo editing guidelines. \\
Original image provided
  & N & Y & Y \\
Chat link provided
  & Y & N & Y \\
\bottomrule
\end{tabularx}
\end{table*}

\begingroup
\small
\setlength{\tabcolsep}{4pt}
\setlength{\stimulusTableWidth}{%
    \dimexpr\textwidth-10\tabcolsep\relax}
\renewcommand{\arraystretch}{1.15}
\setlength{\LTleft}{0pt}
\setlength{\LTright}{0pt}

\begin{longtable}{
    @{}
    >{\raggedright\arraybackslash}p{0.07\stimulusTableWidth}
    >{\raggedright\arraybackslash}p{0.11\stimulusTableWidth}
    >{\raggedright\arraybackslash}p{0.13\stimulusTableWidth}
    >{\raggedright\arraybackslash}p{0.20\stimulusTableWidth}
    >{\raggedright\arraybackslash}p{0.37\stimulusTableWidth}
    >{\raggedright\arraybackslash}p{0.12\stimulusTableWidth}
    @{}
}

\caption{Viewer task stimuli and disclosure information.}
\label{tab:viewer-stimuli}\\
\toprule
Post & Creator disclosure & Detector confidence
& Credited AI technique & Purpose & Model / chat \\
\midrule
\endfirsthead

\multicolumn{6}{@{}l}{%
    \textit{Viewer task stimuli (continued)}}\\
\toprule
Post & Creator disclosure & Detector confidence
& Credited AI technique & Purpose & Model / chat \\
\midrule
\endhead

\midrule
\multicolumn{6}{r@{}}{\textit{Continued on the next page}}\\
\endfoot

\bottomrule
\endlastfoot

\multicolumn{6}{@{}l@{}}{%
    \textbf{Fully AI-generated images}}\\*
\multicolumn{6}{@{}c@{}}{%
    \includegraphics[width=0.98\textwidth]{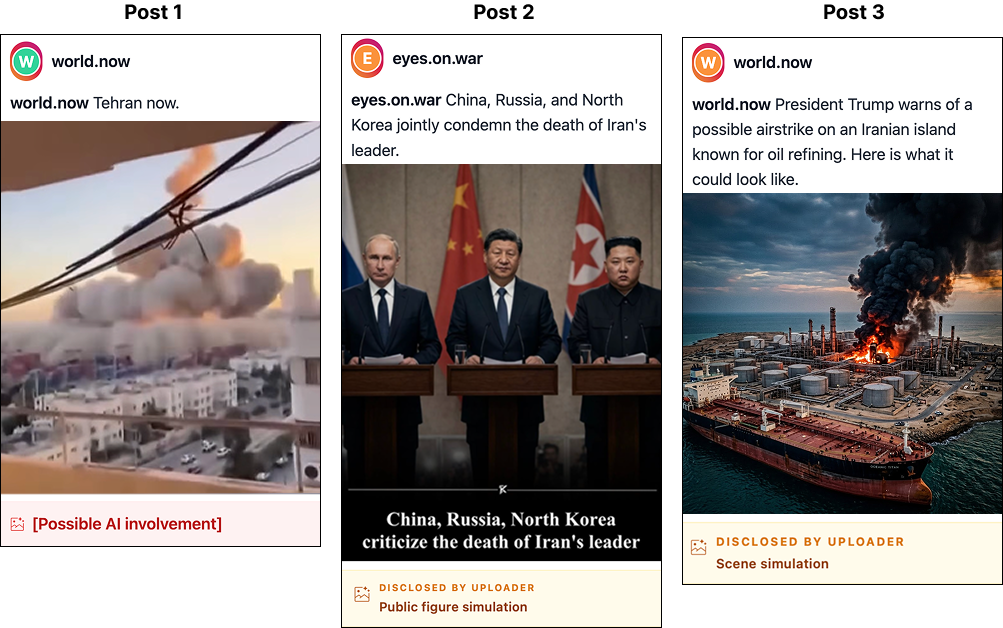}%
}\\[6pt]

Post 1 & N & 54\% & No & N & N \\
\addlinespace[5pt]
Post 2 & Y & 92\%
& Public figure simulation
& To show that the three governments spoke out together.
& Grok AI \\
\addlinespace[5pt]
Post 3 & Y & 98\%
& Scene simulation
& N
& Chat link provided \\

\midrule
\multicolumn{6}{@{}l@{}}{%
    \textbf{Partially AI-generated images}}\\*
\multicolumn{6}{@{}c@{}}{%
    \includegraphics[width=0.98\textwidth]{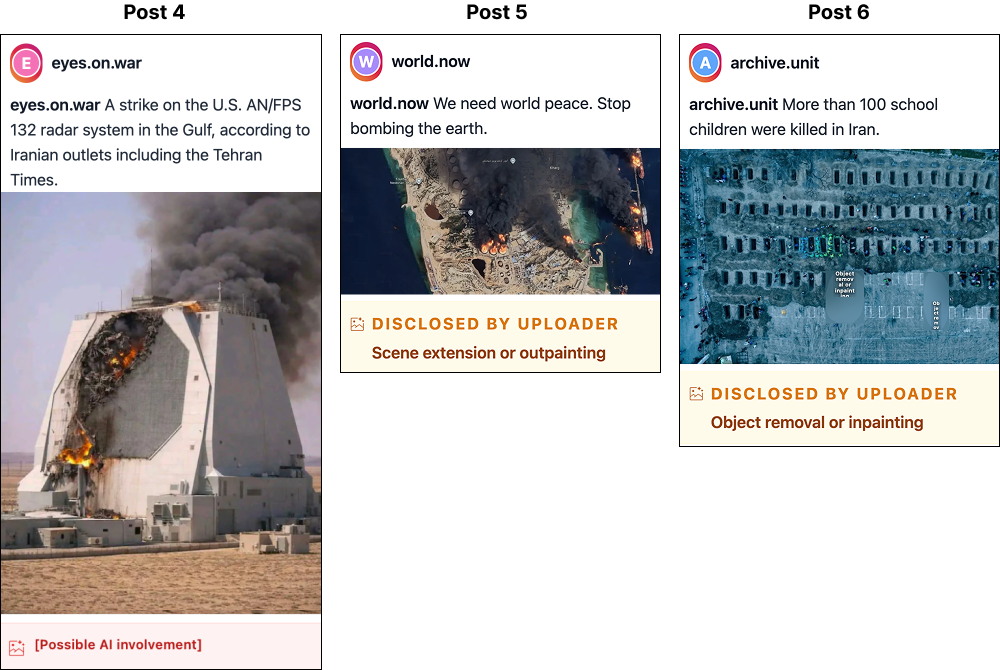}%
}\\[6pt]

Post 4 & N & 51\% & No & N & N \\
\addlinespace[5pt]
Post 5 & Y & 63\%
& Scene extension or outpainting
& To show how the war is unfolding across Iran, Israel, and the US.
& N \\
\addlinespace[5pt]
Post 6 & Y & 51\%
& Object removal or inpainting
& The drone footage was low resolution, so I applied color
grading to sharpen the image. I also removed the yellow
excavator so the graves could be counted more accurately.
& N \\

\midrule
\multicolumn{6}{@{}l@{}}{%
    \textbf{Photographs with no AI involvement}}\\*
\multicolumn{6}{@{}c@{}}{%
    \includegraphics[width=0.98\textwidth]{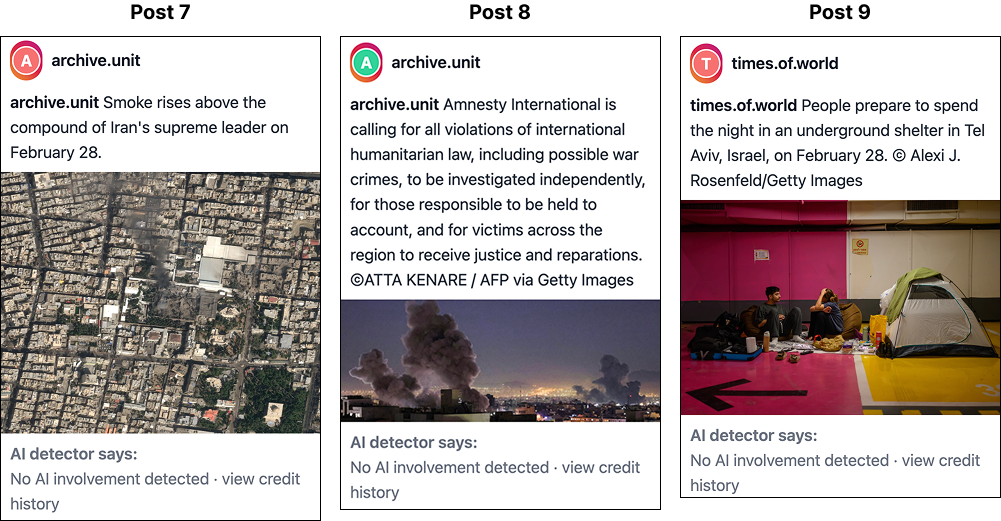}%
}\\[6pt]
\caption{For posts containing original photographs, the detector classified the images as non-AI-generated, and no source images or AI credits were attached.}

\end{longtable}
\endgroup

\subsubsection{Viewer Task}

\paragraph{Study guideline.}
\begin{quote}
\itshape
You are the viewer of an Instagram feed. In your feed, you will see nine
posts covering a current event from accounts you follow. Some used AI,
some did not. You can view the AI usage credit log for each post. You can
like, share, comment on posts, and visit an account to see its content,
follow, unfollow, or share the account. Browse for at least three minutes
and complete the required task list before starting the survey. Screenshot
capture is optional.
\end{quote}

\subsection{Post-Survey}
\label{app:user-study-post}

\subsubsection{Creator Questions}

\textbf{Usability.}
For each statement below, indicate your agreement.
\begin{enumerate}
    \item I think that I would like to use the crediting tool frequently.
    \item I found the crediting tool unnecessarily complex.
    \item I thought the crediting tool was easy to use.
    \item I think that I would need the support of a technical person to be able to use the crediting tool.
    \item I found the various functions in the crediting tool were well integrated.
    \item I thought there was too much inconsistency in the crediting tool.
    \item I would imagine that most people would learn to use the crediting tool very quickly.
    \item I found the crediting tool very cumbersome to use.
    \item I felt confident using the crediting tool.
    \item I needed to learn a lot of things before I could get going with the crediting tool.
\end{enumerate}

\textbf{Clarity of crediting fields.}
For each field below, rate how clear its purpose was while you were crediting the three images.
\begin{enumerate}
    \item The toggle button for declaring AI involvement
    \item The text description field for disclosing purpose
    \item The model name field
    \item The conversation log or chat link field
    \item The hand drawn region overlay for marking where AI was used
\end{enumerate}

\textbf{Attention check}
\begin{enumerate}
    \item To show that you are reading carefully, please select ``Disagree'' for this statement.
\end{enumerate}

\textbf{Representation, effort, disclosure concerns, and adoption.}
For each statement below, indicate your agreement.
\begin{enumerate}
    \item The two images you credited involved different degrees of AI use, one fully generated and one only color corrected using AI. The system allowed me to represent this difference accurately.
    \item A viewer looking at my credited posts would be able to tell the difference between the two images in how AI was used.
    \item Completing the crediting process was time consuming
    \item I believe the effort required to credit AI use adds credibility to my disclosure.
    \item I felt comfortable sharing my prompt and conversation log with the AI model as part of the credit.
    \item I would be concerned about others reusing my prompt if my conversation log were made public.
    \item Knowing that I would need to disclose AI use in this much detail could discourage me from using AI in future content.
    \item The disclosure requirement felt proportionate to the platform’s need for transparency.
    \item Seeing how AI use was presented to viewers helped me understand how to describe my own use of AI.
    \item I would use a system like this in my actual content creation workflow.
    \item I would recommend a system like this to other creators.
\end{enumerate}

\textbf{Trust and open-ended feedback}
\begin{enumerate}
    \item After this task, how much do you trust that content credited with this system accurately represents how it was made?
    \item How easy or difficult would it be for someone to convince others that a real photo or video was AI-generated?
    \item Thinking about the crediting process, what did you find helpful, confusing, or difficult, and why? Please describe a specific example from your experience. If nothing stood out, please say so.
    \item [Optional] What would you change about how AI use is credited?
\end{enumerate}

\subsubsection{Viewer Questions}

\textbf{Usability.}
For each statement below, indicate your agreement.
\begin{enumerate}
    \item I think that I would like to use the credit system in this feed frequently.
    \item I thought the credit system in this feed was easy to use.
    \item I think that I would need the additional support to be able to use to view the credit information.
    \item I found the various functions in the credit system in this feed were well integrated.
    \item I imagine that most people would learn to use the credit system in this feed very quickly.
    \item I felt confident using the credit system in this feed.
    \item I felt viewing the credit system in this feed took too much time.
\end{enumerate}

\textbf{AI recognition}
\begin{enumerate}
    \item Which of the nine posts in the feed did you think were made entirely or partly using AI? Select all that apply.
    \item How confident are you in that estimate?
\end{enumerate}

\textbf{Comprehension.}
For each statement below, indicate your agreement.
\begin{enumerate}
    \item The credit information displayed for each post was easy to understand.
    \item I understood the difference between content that was self declared by the creator and content that was flagged through automated detection.
\end{enumerate}

\textbf{Content credibility and sharing}
\begin{enumerate}
    \item How credible did you find posts with this fully AI generated image?
    \item How likely would you be to share a post like this?
    \item How credible did you find posts with this partially AI generated image?
    \item How likely would you be to share a post like this?
    \item How credible did you find posts with this non-AI generated image?
    \item How likely would you be to share a post like this?
\end{enumerate}

\textbf{Attention check}
\begin{enumerate}
    \item To show that you are reading carefully, please select ``Disagree'' for this statement.
\end{enumerate}

\textbf{Perceived transparency}
\begin{enumerate}
    \item The platform gave me enough information to judge whether a post involved AI.
    \item I felt the credit information was trustworthy.
    \item I felt I could trust the credit information to be accurate.
\end{enumerate}

\textbf{Regional information}
\begin{enumerate}
    \item I could identify which parts of an image were indicated as AI-generated or altered.
    \item I could understand how AI had been used to generate or alter the image.
    \item The information helped me decide how much to trust an image.
    \item The information helped me decide whether to engage with a post.
    \item I would use this feature to filter AI-generated or altered content on social media platforms I regularly use.
\end{enumerate}

\textbf{Reputation and fairness}
\begin{enumerate}
    \item The profile provided enough information for me to assess the creator’s reputation.
    \item The profile provided enough information for me to decide whether to follow this account.
    \item The profile information influenced how much I trusted a post.
\end{enumerate}

\textbf{Trust judgments}
\begin{enumerate}
    \item After browsing the feed, which posts did you trust to accurately show what they claimed to depict? Select all that apply.
    \item How easy or difficult would it be for someone to convince others that a real photo or video was AI-generated?
\end{enumerate}

\textbf{Adoption intention}
\begin{enumerate}
    \item I would use a system like this on the platforms I actually use.
    \item This system would change how I evaluate content I see on social media.
    \item I think this credit and AI detection system can accurately identify AI-generated content.
\end{enumerate}

\textbf{Open-ended feedback}
\begin{enumerate}
    \item Was there any post where the credit information changed your impression of the content? Please describe.
    \item [Optional] What, if anything, was confusing about the credit system?
\end{enumerate}

\section{Study Results: Creator Region Disclosure}
\begin{table}[H]
\centering
\small
\renewcommand{\arraystretch}{1}
\setlength{\tabcolsep}{3pt}
\caption{Three creators' disclosures for the same image.}
\label{tab:disclosure-comparison}

\begin{tabularx}{\linewidth}{
    @{}>{\raggedright\arraybackslash}p{0.13\linewidth}
    *{3}{>{\raggedright\arraybackslash}X}@{}
}
\toprule
Disclosure & C7 & C13 & C14 \\
\midrule

Marked regions &
\begin{minipage}[t]{\linewidth}
    \vspace{0pt}
    \includegraphics[
    width=\linewidth,
    height=38mm,
    keepaspectratio
]{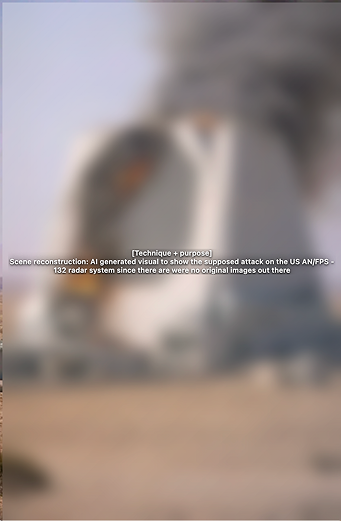}
\end{minipage} &
\begin{minipage}[t]{\linewidth}
    \vspace{0pt}
    \includegraphics[
    width=\linewidth,
    height=38mm,
    keepaspectratio
]{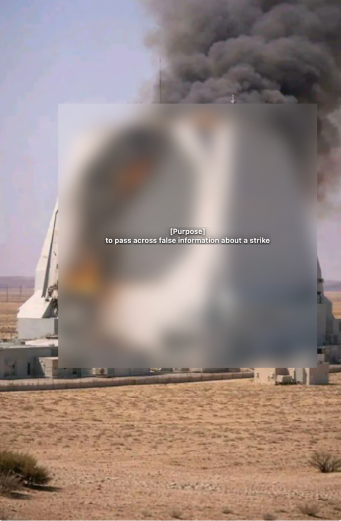}
\end{minipage} &
\begin{minipage}[t]{\linewidth}
    \vspace{0pt}
    \includegraphics[
    width=\linewidth,
    height=38mm,
    keepaspectratio
]{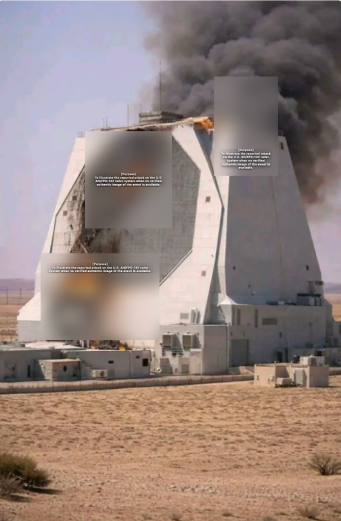}
\end{minipage} \\
\addlinespace[6pt]

AI technique
& 
& Scene simulation
& \\
\addlinespace[6pt]

Purpose
& AI generated visual to show the supposed attack on the
US AN/FPS - 132 radar system since there are were no
original images out there
& to pass across false information about a strike
& To illustrate the reported attack on the U.S. AN/FPS-132
radar system when no verified authentic image of the event
is available. \\
\addlinespace[6pt]

Model
& Unknown
& N
& Google Gemini image generation \\
\addlinespace[6pt]

Chat log
& N
& N
& Generate a realistic, news-style image depicting a
hypothetical Iranian strike on the U.S. AN/FPS-132 radar
installation in the Gulf, showing damage and smoke around
the radar facility. The image is illustrative, not verified
footage of an actual attack. \\

\bottomrule
\end{tabularx}
\end{table}